\documentclass[11pt, a4paper, logo, copyright, nonumbering]{map}
\usepackage{CJKutf8}
\usepackage{xargs} 
\usepackage{todonotes}
\usepackage{amsmath}
\usepackage{dsfont}

\usepackage{longtable}

\usepackage{svg}
\usepackage{fontawesome}
\usepackage{array}
\usepackage{tabularx}
\usepackage{latexsym}
\usepackage{graphicx}
\usepackage[utf8]{inputenc} 
\usepackage[utf8]{inputenc} 
\usepackage{amssymb} 
\usepackage{url}            
\usepackage{amsfonts}       
\usepackage{nicefrac}       
\usepackage{microtype}      
\usepackage{xspace}

\usepackage{listings}
\usepackage{makecell}
\usepackage{inconsolata}
\usepackage{multicol}
\usepackage{amstext}

\definecolor{darkblue}{RGB}{84, 112, 198}

\definecolor{lightblue}{rgb}{0.85, 0.95, 1.0}    
\definecolor{lightgreen}{rgb}{0.90, 1.0, 0.90}    
\definecolor{lightorange}{rgb}{1.0, 0.95, 0.85}   
\definecolor{lightpurple}{rgb}{0.95, 0.90, 1.0}   
\definecolor{lightgray}{rgb}{0.97, 0.97, 0.97}    

\usepackage{tikz}
\usetikzlibrary{calc}
\definecolor{battery-empty}{rgb}{0.9, 0.9, 0.9}
\newcommand{\difficultybar}[1]{%
  \begin{tikzpicture}[baseline, scale=0.5, every node/.style={scale=0.8}]
    \foreach \i in {1,2,3,4,5} {
      \ifnum\i>#1
        \draw[fill=battery-empty] (\i*0.5-0.5, 0) rectangle (\i*0.5, 0.25);
      \else
        \pgfmathsetmacro{\colorlevel}{80 - 12*(\i)} 
        \edef\x{\noexpand\draw[fill=blue!\colorlevel!white, opacity=0.9] (\i*0.5-0.5, 0) rectangle (\i*0.5, 0.25);}
        \x
        \draw[blue!50!black] (\i*0.5-0.5, 0) rectangle (\i*0.5, 0.25);
      \fi
    }
    \fill[battery-empty!70] (2.5, 0.08) rectangle (2.6, 0.17);
    \draw[battery-empty!70!black] (2.5, 0.08) rectangle (2.6, 0.17);
  \end{tikzpicture}%
}
\renewcommand{\arraystretch}{0.96}

\definecolor{hidden-draw}{RGB}{20,68,106}
\definecolor{hidden-pink}{RGB}{255,245,247}
\usepackage[edges]{forest}

\usepackage{bm}

\usepackage{natbib}
\usepackage{CJKutf8}
\usepackage{xargs}
\usepackage{todonotes}
\usepackage{multirow}
\usepackage{cleveref}
\usepackage{amsmath}
\usepackage{dsfont}
\usepackage{subcaption}
\usepackage{url}
\usepackage{svg}
\usepackage{fontawesome}
\usepackage{xcolor}
\usepackage{float}

\usepackage[utf8]{inputenc}
\usepackage{booktabs}
\usepackage{graphicx}
\usepackage{array}
\usepackage{tabularx}
\usepackage{xcolor,colortbl}  
\definecolor{boxcolor}{HTML}{d92523} 
\definecolor{bulbcolor}{HTML}{e3b87f} 
\usepackage{url}
\usepackage{ragged2e}  
\usepackage{makecell} 

\usepackage{hyperref}       
\usepackage{url}            
\usepackage{booktabs}       
\usepackage{amsfonts}       
\usepackage{nicefrac}       
\usepackage{microtype}      
\usepackage{xcolor}         
\usepackage{graphicx}
\usepackage{longtable}
\usepackage{array}
\usepackage{pbox}
\usepackage{amsmath}
\usepackage{amssymb}
\usepackage{tabularx}
\usepackage{multirow}
\usepackage{wrapfig}
\usepackage{pifont}
\usepackage{algorithm}
\usepackage{algorithmic}
\usepackage{lipsum}
\usepackage{subcaption}
\usepackage{enumitem}
\usepackage{tikz}
\usepackage{xstring}

\usepackage{color-edits}

\usepackage{booktabs}
\usepackage{multirow}
\usepackage[table]{xcolor}
\usepackage{tabularx}
\usepackage{graphicx}
\usepackage{hyperref}

\usepackage{parskip} 

\usepackage{threeparttable}

\newcommand{\modelname}{InCoder-32B-Thinking} 
\definecolor{rliableolive}{HTML}{BBCC33}
\definecolor{rliableblue}{HTML}{77AADD}
\definecolor{rliablered}{HTML}{f63c44}

\definecolor{rliableolive}{HTML}{BBCC33}
\definecolor{rliableblue}{HTML}{77AADD}
\definecolor{rliablered}{HTML}{f63c44}

\usepackage[most,skins,theorems]{tcolorbox}
\tcbuselibrary{skins,breakable,fitting,hooks}  
\tcbset{
  aibox/.style={
    width=\linewidth,
    top=7pt,
    bottom=2pt,
    colback=rliablered!18!white,
    colframe=black,
    colbacktitle=black,
    enhanced,
    center,
    attach boxed title to top left={yshift=-0.1in,xshift=0.15in},
    boxed title style={boxrule=0pt,colframe=white,},
  }
}
\tcbset{
  aibox2/.style={
    width=\linewidth,
    top=7pt,
    bottom=2pt,
    colback=green!18!white,
    colframe=black,
    colbacktitle=black,
    enhanced,
    center,
    attach boxed title to top left={yshift=-0.1in,xshift=0.15in},
    boxed title style={boxrule=0pt,colframe=white,},
  }
}
\newtcolorbox{AIbox}[2][]{aibox,title=#2,#1}
\newtcolorbox{AIbox2}[2][]{aibox2,title=#2,#1}

\hypersetup{
    colorlinks=true,            
    linkcolor=blue,            
    filecolor=magenta,          
    urlcolor=cyan,             
    citecolor=purple,            
    pdftitle={Overleaf Example},
    pdfpagemode=FullScreen,
}

\definecolor{iquestblue}{HTML}{173C7F}
\definecolor{iquestazure}{HTML}{528FCC}

\renewcommand{\today}{}
\newcommandx{\info}[2][1=]{\todo[linecolor=red,backgroundcolor=red!25,bordercolor=red,#1]{#2}}

\title{
\vspace{-0.2in}
\centering \fontsize{15pt}{16pt}\selectfont
\modelname{}: Industrial Code World Model for Thinking
\vspace{-0.2in}
}

\author{
Jian Yang\textsuperscript{1}, 
Wei Zhang\textsuperscript{1}, 
Jiajun Wu\textsuperscript{1}, 
Junhang Cheng\textsuperscript{1}, 
Tuney Zheng\textsuperscript{2},
Fanglin Xu\textsuperscript{2},
Weicheng Gu\textsuperscript{1},
Lin Jing\textsuperscript{2},
Yaxin Du\textsuperscript{3},
Joseph Li\textsuperscript{4},
Yizhi Li\textsuperscript{5},
Yan Xing\textsuperscript{2},
Chuan Hao\textsuperscript{2},
Ran Tao\textsuperscript{2},
Ruihao Gong\textsuperscript{1},
Aishan Liu\textsuperscript{1},
Zhoujun Li\textsuperscript{1},
Mingjie Tang\textsuperscript{7},
Chenghua Lin\textsuperscript{5},
Siheng Chen\textsuperscript{3},
Wayne Xin Zhao\textsuperscript{8$\dagger$},
Xianglong Liu\textsuperscript{1$\dagger$}, %
Ming Zhou\textsuperscript{9$\dagger$},
Bryan Dai\textsuperscript{2},
Weifeng Lv\textsuperscript{1}\\
\textsuperscript{1}Beihang University
\textsuperscript{2}IQuest Research
\textsuperscript{3}Shanghai Jiao Tong University
\textsuperscript{4}ELLIS
\textsuperscript{5}University of Manchester
\textsuperscript{7}Sichuan University
\textsuperscript{8}Gaoling School of Artificial Intelligence, Renmin University of China
\textsuperscript{9}Langboat
\\
\textsuperscript{$\dagger$}Corresponding Authors. Email: \texttt{\{jiayang\}@buaa.edu.cn}

{
\includegraphics[height=1em]{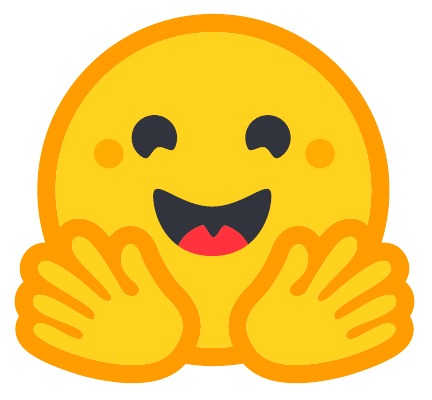}\;
HuggingFace: \url{https://huggingface.co/Multilingual-Multimodal-NLP/IndustrialCoder}
}\\

{
\includegraphics[height=1em]{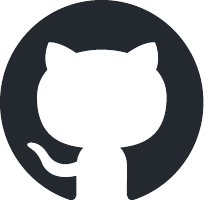}\;
GitHub: \url{https://github.com/CSJianYang/Industrial-Coder}
}
\vspace{-30pt}
}

\begin{abstract}
\vspace{-0.75em} 
Industrial software development across chip design, GPU optimization, and embedded systems lacks expert reasoning traces showing how engineers reason about hardware constraints and timing semantics. In this work, we propose \modelname{}, trained on the data from the Error-driven Chain-of-Thought (ECoT) synthesis framework with an industrial code world model (ICWM) to generate reasoning traces. Specifically, ECoT generates reasoning chains by synthesizing the thinking content from multi-turn dialogue with environmental error feedback, explicitly modeling the error-correction process. ICWM is trained on domain-specific execution traces from Verilog simulation, GPU profiling, etc., learns the causal dynamics of how code affects hardware behavior, and enables self-verification by predicting execution outcomes before actual compilation. All synthesized reasoning traces are validated through domain toolchains, creating training data matching the natural reasoning depth distribution of industrial tasks. Evaluation on 14 general (81.3\% on LiveCodeBench v5) and 9 industrial benchmarks (84.0\% in CAD-Coder and 38.0\% on KernelBench) shows \modelname{} achieves top-tier open-source results across all domains. 
\end{abstract}

\begin{document}

\maketitle

\let\oldthefootnote\thefootnote

\begin{figure*}[h!]
    \centering
\includegraphics[width=0.8\textwidth]{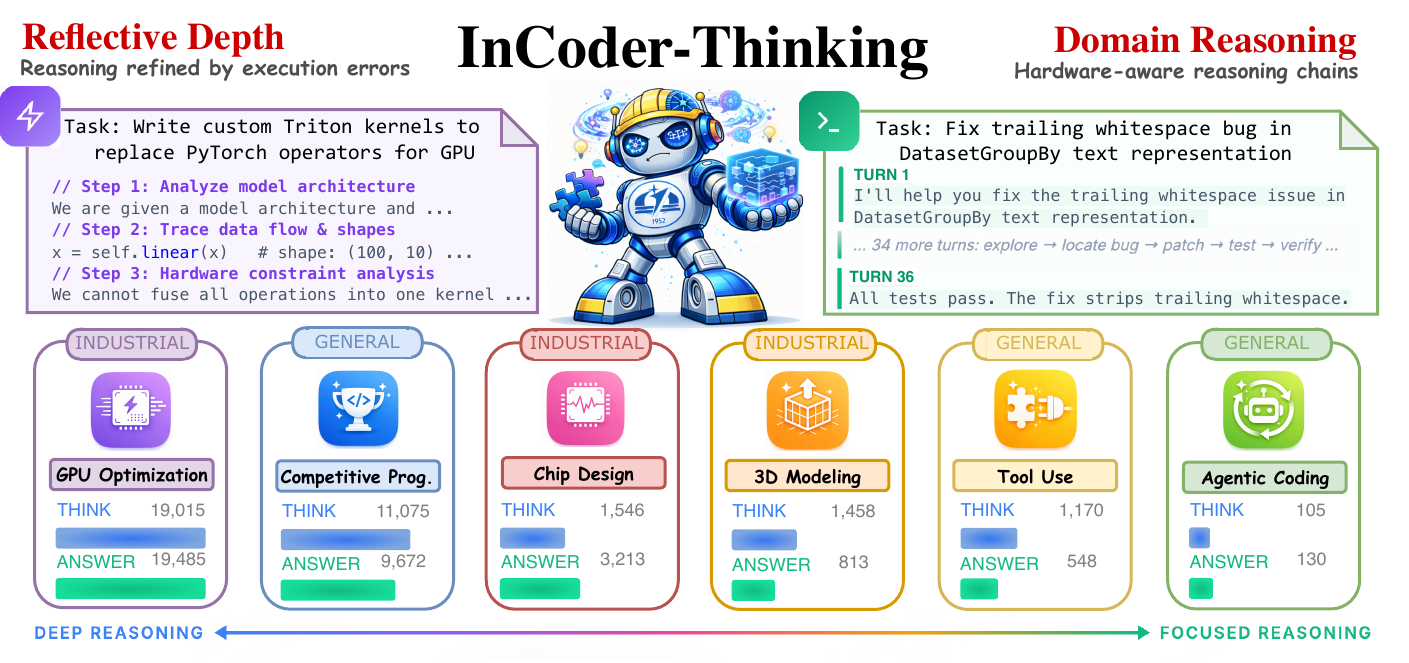}
    \vspace{-5pt}
    \caption{Overview of \modelname, a coder for general and industrial code intelligence with thinking capability, which supports tasks from general coding to industrial tasks such as GPU optimization, chip design, and 3D modeling.}
    \label{fig:tease_perf}
\end{figure*}

\begin{figure*}[h!]
    \centering
    \includegraphics[width=\textwidth]{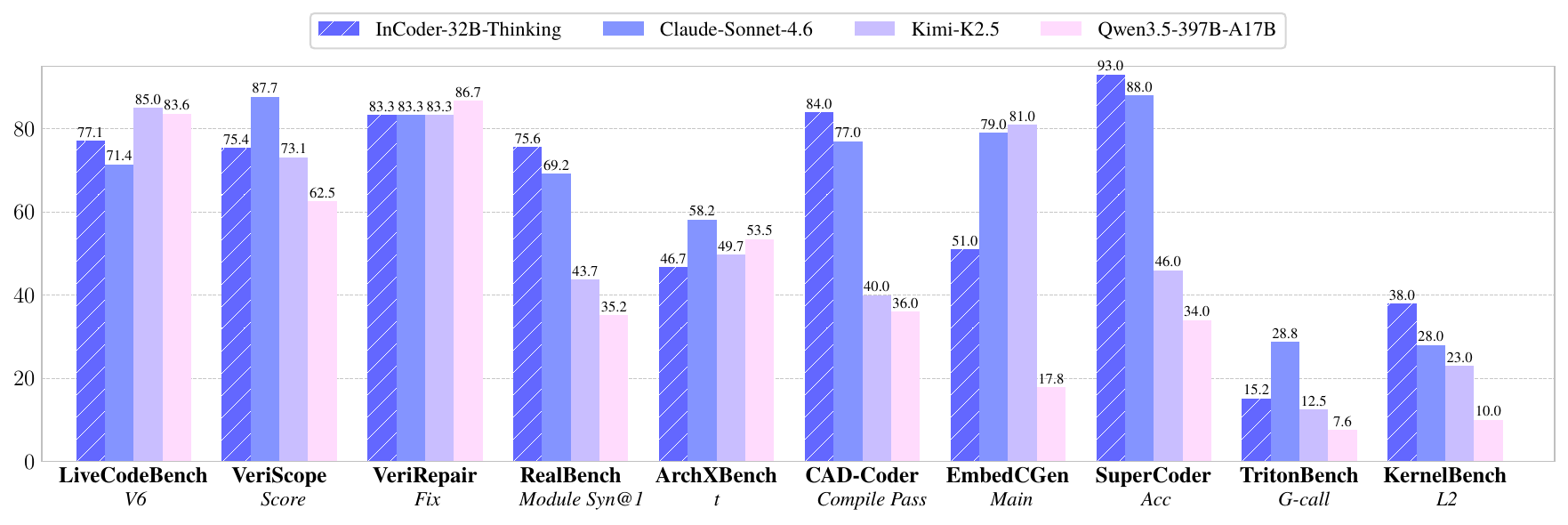}
    \vspace{-5pt}
    \caption{The performance of \modelname{} on code benchmarks.}
    \label{fig:benchmark_compare}
\end{figure*}


\begin{figure*}[t!]
    \centering
    \includegraphics[width=\textwidth]{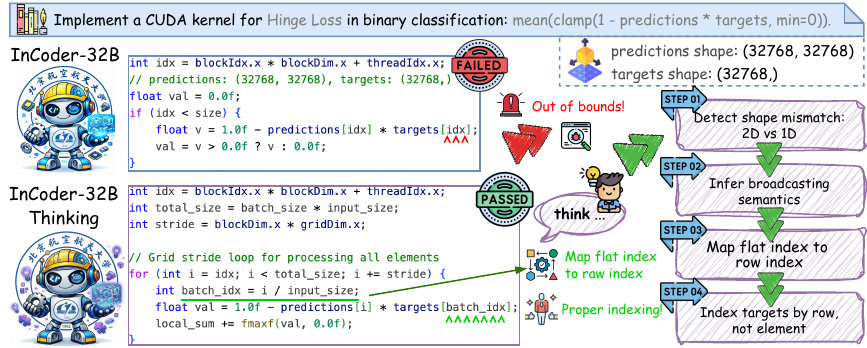}
    \caption{Comparison of CUDA kernel implementations for Hinge Loss. Through reasoning, \modelname{} identifies the shape mismatch between the 2D predictions and 1D targets tensors, correctly mapping flat indices to row indices for proper broadcasting. The non-thinking model indexes both tensors with the same flat index, causing out-of-bounds error.}
    \label{fig:intro_example}
\end{figure*}

\section{Introduction}
Large language models (LLMs) have achieved remarkable advances in code intelligence~\cite{yang2025codesurvey}. Recent LLMs, such as DeepSeek-V3.2~\cite{liu2025deepseekv32} and Claude-4.6~\cite{claude46} have demonstrated strong performance in software tasks~\cite{swebench,swe-bench-live,swebenchmultilingual,tau2bench,servers25mcp}. Further, recent LLMs~\cite{incoder_32b} and benchmarks~\cite{agents4plc,jin2025realbench} highlight the growing importance and complexity of industrial code generation scenarios.
The thinking models exemplified by OpenAI o-series~\cite{o3-o4} and DeepSeek-R1~\cite{deepseekai2025deepseekr1} that generate extended chain-of-thought reasoning traces to break down problems into verifiable steps.

While thinking models excel at reasoning and world models at predicting dynamics, their integration for industrial code generation remains unexplored. Thinking models simulate execution through deliberation, while code world models~\cite{cwm} forecast outcomes and inject feedback into reasoning. Tasks such as CUDA optimization~\cite{kernelbench}, Verilog description~\cite{verilog_eval}, firmware programming~\cite{firm_embedded_programming_llm}, and compiler optimization~\cite{compiler_optimization_llm} require reasoning about specialized semantics and hardware constraints absent from web-scale corpora. Verification depends on complex toolchains whose behavior must be learned for domain-specific feedback injection. Benchmarks show leading models achieve limited success on Triton generation~\cite{tritonbench} and Verilog checking~\cite{jin2025realbench} despite strong general performance. This gap suggests industrial code generation requires integrating reasoning with learned toolchain dynamics.

To address these challenges, we propose \modelname{}, trained through two synergistic components: \textbf{Error-driven Chain-of-Thought (ECoT) synthesis} that generates reasoning traces by explicitly modeling error-correction processes, and an \textbf{Industrial Code World Model (ICWM)} that learns causal dynamics between code and hardware behavior from domain toolchain executions. Industrial code reasoning needs environment grounding to predict hardware effects of code execution and validate via a learned simulator before actual execution. The ECoT framework generates reasoning traces by contrasting failed attempts with correct solutions, mimicking the diagnostic processes of engineers. While the ICWM simulates environmental feedback via domain-specific execution traces (Verilog logs, GPU profiles, etc), enabling self-verification, efficient exploration without toolchain execution, and synthetic failure scenario generation. \modelname{} is trained on the data from ECoT synthesis and the data from teacher-student distillation, where complex cases accumulate multi-step error-correction and reasoning traces. 

Extensive evaluations on general and industrial code benchmarks demonstrate that \modelname{} combines broad coding competence with specialized industrial capabilities. \modelname{} achieves 70.4\% on SWE-bench Verified~\cite{swebench}, 81.3\% on LiveCodeBench\_v5~\cite{jain2024livecodebench}, and 63.9\% on BFCL~\cite{patil2025bfcl}, competitive with leading models of comparable or larger scale. Compared to its instruct counterpart, \modelname{} achieves comparable performance on SWE-bench Verified, while improving by 28.0\% on LiveCodeBench, demonstrating the effectiveness of continued pretraining on code data. On industrial benchmarks, \modelname{} establishes the strongest open-source results across all evaluated domains, including chip design~\cite{guan2025cad}, GPU kernel optimization~\cite{kernelbench}, embedded systems~\cite{xu2025embedagent}, compiler optimization~\cite{tritonbench}, and 3D modeling~\cite{guan2025cadcodertexttocadgenerationchainofthought}. Our contributions are:
\begin{itemize}
\item \textbf{Error-driven thinking synthesis for industrial code.} We propose Error-driven Chain-of-Thought (ECoT) synthesis that generates reasoning traces by explicitly modeling the error-correction process through contrasting incorrect attempts and their environmental feedback with correct solutions, capturing the iterative refinement patterns that define industrial engineering expertise.

\item \textbf{Industrial Code World Model for industrial environments.} We develop the first world model for industrial code environments, trained on domain-specific execution traces (Verilog simulation, GPU profiling, compiler diagnostics, embedded system logs) to learn causal dynamics between code modifications and hardware behavior, enabling self-verification, efficient exploration, and synthetic trace generation without expensive toolchain execution.

\item \textbf{Industrial code thinking model.} \modelname{} integrates ECoT-synthesized reasoning traces from world model predictions to achieve top-tier open-source results across different coding domains, demonstrating that thinking models grounded in learned environment dynamics are key to real-world industrial code intelligence.
\end{itemize}

\section{Error-driven thinking synthesis from Industrial Code World Model}
\label{sec:pipeline}
\begin{figure*}[ht!]
    \centering
    \includegraphics[width=\textwidth]{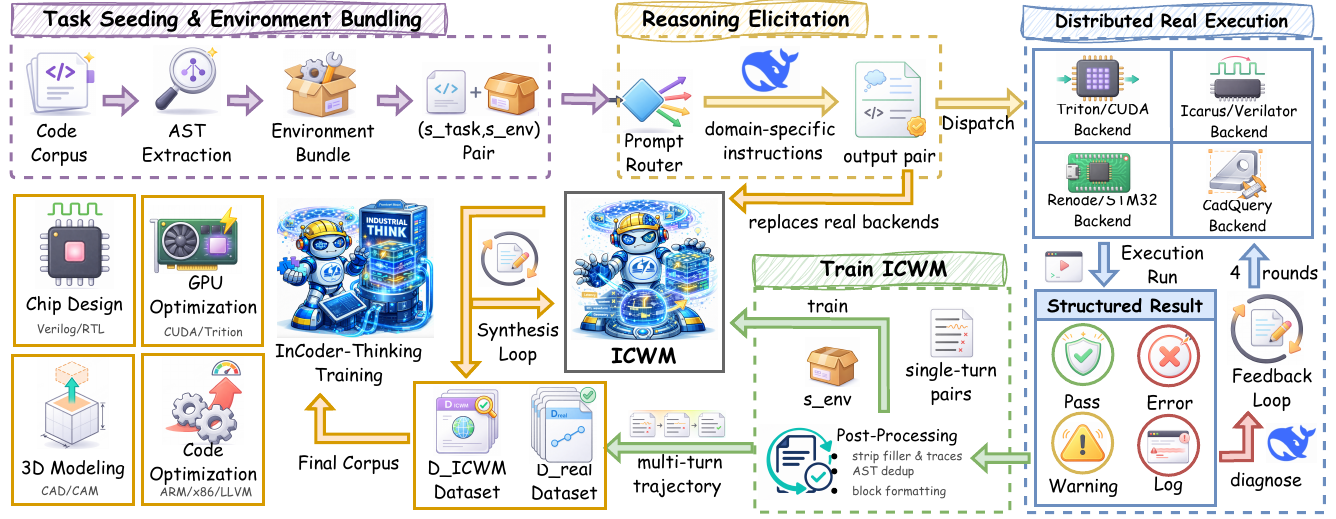}
    \caption{Overview of the data engine pipeline.
    \textbf{Left:} task seeds and environment bundles are passed through an elicitation, execution, feedback loop against real backends, producing multi-turn trajectories $\mathcal{D}_{\text{real}}$.
    \textbf{Right:} $\mathcal{D}_{\text{real}}$ trains an \textsc{ICWM} that simulates the real backends during large-scale synthesis; periodic audits keep the world model calibrated across iterations.}
    \label{fig:pipeline_flow}
\end{figure*}

In this section, we first collect the query and multi-turn responses with the real execution feedback. Then, we train the industrial code world model (ICWM) on the collected data, which enables large-scale simulation without repeated access to real backends. Finally, we use the ICWM to simulate the multi-turn execution feedback and synthesize the thinking content from multi-turn dialogues, where the thinking content finds the final correct answer through multiple failed attempts. The pipeline operates in two phases (\autoref{fig:pipeline_flow}):

\begin{enumerate}
\item \textbf{Grounded collection (\autoref{subsec:pipe_seed} and \autoref{subsec:pipe_traj}).} A generator produces code, executes it against real toolchains and simulators, records structured feedback, and iterates for up to $K$ correction rounds. The resulting multi-turn trajectories pair reasoning traces with concrete execution outcomes and directly enter the training corpus $\mathcal{D}_{\text{real}}$.
\item \textbf{ICWM driven amplification (\autoref{subsec:icwm}).} The collected trajectories also train an ICWM that predicts execution outcomes without invoking real backends. The \textsc{ICWM} then serves as a fast proxy environment for large-scale trajectory synthesis, with periodic real execution audits that keep the world model calibrated.
\end{enumerate}


\subsection{Task Seeding and Environment Bundling}
\label{subsec:pipe_seed}
We reuse the data from InCoder-32B~\cite{incoder_32b}, including queries, unit tests, and environments. Each task seed is packaged with its required environmental context: Verilog modules are bundled with testbenches and Yosys synthesis scripts; STM32 firmware snippets are coupled with the corresponding memory layout, CMSIS headers, and linker scripts. The resulting pair $\langle s_{\text{task}},\; s_{\text{env}} \rangle$ provides a fully specified, reproducible execution context for every downstream step.

\subsection{Execution Grounded Trajectory Synthesis}
\label{subsec:pipe_traj}

Given a task seed and its environment bundle, a frontier model is prompted to produce a reasoning trace and candidate code. Before generation begins, a lightweight prompt router inspects the environment bundle and selects domain-specific instructions for the generator. For example, GPU kernel tasks require the generator to reason about warp divergence and shared memory budgets; RTL tasks require reasoning about combinational path depth and clock domain crossings; CAD tasks about wall thickness and manifold validity; firmware tasks about peripheral register sequencing.  With these instructions in place, the generator DeepSeek-V3.2 outputs an initial pair of reasoning content and code $(r^{(0)}, c^{(0)})$.

\paragraph{Distributed execution.}
The candidate code $c^{(0)}$ is sent to real execution backends that match its domain: Triton/CUDA for GPU kernels, Renode for microcontroller firmware, CadQuery for CAD geometry, and Yosys/Icarus for RTL.  Each backend returns a structured result containing (i)~an outcome label such as \textsc{Pass}, \textsc{Compilation\_Error}, or \textsc{Memory\_Fault}, and (ii)~the associated diagnostic log.

\paragraph{Synthesizing reasoning content from multi-turn interaction.}
When execution produces an error, the concrete output, compilation logs, hardware exceptions, geometric diff reports, is packaged as an observation
$o^{(k)}$ and fed back to the generator, which must diagnose the fault and
produce a revised trace and implementation.  We allow up to $K=4$
correction rounds, yielding a multi-turn trajectory:
\begin{equation}
  \tau = \bigl[\,
    \langle s_{\text{init}},\, r^{(0)},\, c^{(0)} \rangle
    \;\xrightarrow{}\;
    \langle r^{(1)},\, c^{(1)} \rangle
    \;\cdots\;
    \xrightarrow{}\;
    \langle r^{(k)},\, c^{(k)} \rangle
  \,\bigr]
\end{equation}where $s_{\text{init}}$ is the initial task description together with its environment bundle, $r^{(k)}$ is the reasoning content at turn~$k$, $c^{(k)}$ is the corresponding candidate code, and each arrow represents one round of execution feedback followed by the generator's revision. A trajectory terminates when the code passes all checks or when the maximum number of turns~$K$ is reached.

Both successful and unsuccessful intermediate turns are retained so that training data captures common failure modes alongside the reasoning steps that resolve them. The full multi-turn trajectories in $\mathcal{D}_{\text{real}}$ train the code model, while the individual single-turn pairs extracted from them train the world model described next.

\subsection{ICWM Driven Data Amplification}
\label{subsec:icwm}
Real execution provides reliable supervision but is expensive for complex industrial tasks because each interaction requires invoking domain-specific toolchains and maintaining a robust execution stack. To improve scalability, we train a learned proxy, denoted as \textsc{ICWM}, to approximate backend feedback conditioned on the environment and candidate code.

\paragraph{Preliminaries.}
We define the \emph{world} in an industrial coding task as the executable environment that determines the consequence of a code revision. Concretely, an environment bundle $s_{\text{env}}$ contains the task-specific artifacts required for execution and verification, such as testbenches, simulators, compiler settings, configuration files, memory layouts, and domain constraints. At turn $k$, the generator proposes a candidate program $c^{(k)}$, which acts on the environment. Executing $c^{(k)}$ under $s_{\text{env}}$ yields observable feedback $o^{(k)}$, including an outcome label, diagnostic logs, and task-specific outputs.

\paragraph{Definition of ICWM.}
The \textsc{ICWM} is a language model that takes the task environment and a piece of candidate code as input, and predicts what a real backend would return: the ICWM can be understood as a learned proxy of industrial execution backends. Given an environment bundle $s_env$ and a candidate program $c$, it predicts the observable feedback o that would be returned by real toolchains, including execution status, diagnostic logs, and task-specific outputs.

\begin{equation}
  \textsc{ICWM}_\theta:\;
  (s_{\text{env}},\; c^{(k)})
  \;\longmapsto\;
  \hat{o}^{(k)}
\end{equation}
where the predicted output $\hat{o}^{(k)}$ includes an outcome label
(\textsc{Pass}, \textsc{Compilation\_Error}, etc.), a diagnostic
message, and, when applicable, numerical outputs or diff
summaries.  The model is trained on every real execution turn in
$\mathcal{D}_{\text{real}}$.  To handle the differences across
domains, we prepend a domain tag to each input and use domain specific output templates, so that a single model can serve all four verticals.

\paragraph{ICWM driven synthesis loop.}
Once trained, the \textsc{ICWM} replaces real backends in the
feedback loop described in \autoref{subsec:pipe_traj}:
\begin{equation}
  \tau = \bigl[\,
    \langle s_{\text{init}},\, r^{(0)},\, c^{(0)} \rangle
    \;\xrightarrow{\;\textsc{ICWM}_\theta\;}\;
    \langle r^{(1)},\, c^{(1)} \rangle
    \;\cdots\;
    \xrightarrow{\;\textsc{ICWM}_\theta\;}\;
    \langle r^{(k)},\, c^{(k)} \rangle
  \,\bigr]
\end{equation}where each arrow represents one round of ICWM feedback rather than execution feedback followed by the generator's revision, since each prediction is a single forward pass rather than a real compilation or simulation.
\paragraph{Resulting reasoning corpus.}
The final reasoning training set is
$\mathcal{D} = \mathcal{D}_{\text{real}}
\cup \mathcal{D}_{\text{icwm}}$.
Every trajectory in $\mathcal{D}$ is either produced by real
execution or verified against it, ensuring the code model
learns from grounded feedback rather than hallucinated outputs.

\definecolor{tablegray}{gray}{0.92}
\begin{table*}[t!]
    \centering
    \caption{Performance comparison on general code generation tasks.}
    \label{tab:code_generation_1}
    \resizebox{1.0\textwidth}{!}{
    \begin{tabular}{lr|cccc|cc|c}
    \toprule
        \multirow{2}{*}{\textbf{Model}} & \multirow{2}{*}{\textbf{Size}} & \multicolumn{4}{c}{\textbf{EvalPlus}} & \multicolumn{2}{|c|}{\textbf{BigCodeBench}} & \multirow{2}{*}{\textbf{FullStackBench}} \\
        ~ & ~ & \textbf{HumanEval} & \textbf{HumanEval+} & \textbf{MBPP} & \textbf{MBPP+} & \textbf{Full} & \textbf{Hard} & ~  \\
        \midrule
        \multicolumn{9}{c}{\textbf{6B+ Models}} \\
        \midrule
        DeepSeek-Coder-V2-Lite-Instruct & 2.4/16B & 81.1 & 75.6 & 85.2 & 70.6 & 37.8 & 18.9 & 49.4 \\
        Qwen2.5-Coder-7B-Instruct & 7B & \textbf{87.2} & \textbf{81.7} & 84.7 & 72.2 & 37.8 & 13.5 & 42.2 \\
        Seed-Coder-8B-Instruct & 8B & 81.1 & 75.6 & 86.2 & 73.3 & 44.6 & \textbf{23.6} & \textbf{55.8} \\
        Qwen2.5-Coder-14B-Instruct & 14B & 62.8 & 59.8 & \textbf{88.6} & \textbf{77.2} & \textbf{47.0} & 6.1 & 53.1 \\
        \midrule
        \multicolumn{9}{c}{\textbf{30B+ Models}} \\
        \midrule
        Qwen3-Coder-30B-A3B-Instruct & 3.3/30.5B & 93.9 & 87.2 & 90.7 & 77.2 & 46.9 & 27.7 & 60.9 \\
        Deepseek-V3.2 & 37/671B & 93.9 & 88.4 & 93.4 & 77.2 & 48.1 & 27.0 & 64.9 \\
        Qwen2.5-Coder-32B-Instruct & 32B & 93.3 & 86.6 & 90.2 & 77.8 & 48.0 & 24.3 & 57.4 \\
        Qwen3-235B-A22B-Instruct-2507 & 22/235B & 96.3 & 91.5 & 92.3 & 77.8 & 47.4 & 25.7 & 62.7 \\
        Qwen3-235B-A22B-Thinking-2507 & 22/235B & \textbf{98.8} & \textbf{93.3} & 95.5 & 81.5 & 44.1 & 23.0 & - \\
        Qwen3-Coder-480B-A35B-Instruct & 35/480B & 97.6 & 92.7 & 94.2 & 80.2 & 49.4 & 27.7 & 66.4 \\
        Kimi-Dev-72B & 72B & 93.3 & 86.0 & 79.6 & 68.8 & 45.4 & \textbf{31.8} & 38.6 \\
        Kimi-K2-Instruct-0905 & 32B/1T & 94.5 & 89.6 & 91.8 & 74.1 & \textbf{49.8} & 30.4 & 63.5 \\
        Kimi-K2-Thinking & 32B/1T & 98.2 & 92.7 & \textbf{97.4} & \textbf{82.3} & 46.8 & 28.4 & - \\
        KAT-Dev & 32B & 90.9 & 86.6 & 89.4 & 76.2 & 46.2 & 25.7 & 58.8 \\
        KAT-Dev-72B-Exp & 72B & 88.4 & 81.7 & 85.2 & 69.3 & 48.3 & 26.4 & 52.9 \\
        GLM-4.7 & 32/355B & 87.2 & 79.9 & 90.5 & 75.7 & 45.7 & 26.4 & \textbf{70.2} \\
        InCoder-32B & 32B & 94.5 & 89.6 & 91.8 & 78.3 & \textbf{49.8} & 31.1 & 57.1 \\
        \rowcolor{tablegray} \textbf{\modelname{}} & 32B & 95.1 & 89.6 & 92.1 & 78.3 & 47.4 & 29.1 & 60.8 \\
    \bottomrule
    \end{tabular}
    }
\end{table*}

\definecolor{tablegray}{gray}{0.92}
\begin{table*}[!h]
    \centering
    \caption{Performance comparison on code reasoning (CruxEval, LiveCodeBench), code efficiency (Mercury), and Text2SQL (Bird, Spider) benchmarks.}
    \label{tab:code_reasoning_efficiency_sql}
    \resizebox{1.0\textwidth}{!}{
    \begin{tabular}{lr|cccc|cc|cc}
    \toprule
        \multirow{3}{*}{\textbf{Model}} & \multirow{3}{*}{\textbf{Size}} & \multicolumn{4}{c|}{\textbf{Code Reasoning}} & \multicolumn{2}{c|}{\textbf{Code Efficiency}} & \multicolumn{2}{c}{\textbf{Text2SQL}} \\
        \cmidrule(lr){3-6} \cmidrule(lr){7-8} \cmidrule(lr){9-10}
        ~ & ~ & \multicolumn{2}{c}{\textbf{CruxEval}} & \multicolumn{2}{c|}{\textbf{LiveCodeBench}} & \multicolumn{2}{c|}{\textbf{Mercury}} & \multirow{2}{*}{\textbf{Bird}} & \multirow{2}{*}{\textbf{Spider}} \\
        ~ & ~ & \textbf{Input-COT} & \textbf{Output-COT} & \textbf{V5} & \textbf{V6} & \textbf{Beyond@1} & \textbf{Pass@1} & ~ & ~ \\
        \midrule
        \multicolumn{10}{c}{\textbf{6B+ Models}} \\
        \midrule
        DeepSeek-Coder-V2-Lite-Instruct & 2.4/16B & 57.1 & 56.2 & 13.2 & 19.4 & 76.8 & 91.4 & 41.6 & 72.4 \\
        Qwen2.5-Coder-7B-Instruct & 7B & 66.9 & 66.0 & 14.4 & 18.9 & 69.9 & 84.8 & 53.1 & 79.8 \\
        Seed-Coder-8B-Instruct & 8B & 62.0 & 66.6 & 19.2 & 22.3 & \textbf{78.5} & \textbf{93.8} & 44.7 & 72.7 \\
        Qwen2.5-Coder-14B-Instruct & 14B & \textbf{75.6} & \textbf{79.2} & \textbf{22.8} & \textbf{24.6} & 76.7 & 88.3 & \textbf{59.1} & \textbf{81.3} \\
        \midrule
        \multicolumn{10}{c}{\textbf{30B+ Models}} \\
        \midrule
        Qwen3-Coder-30B-A3B-Instruct & 3.3/30.5B & 76.9 & 80.5 & 43.1 & 36.0 & 81.1 & 95.3 & 59.0 & 80.9 \\
        DeepSeek-v3.2 & 37/671B & 82.1 & 94.2 & - & 83.3 & \textbf{81.6} & \textbf{96.9} & 52.6 & 77.9 \\
        Qwen2.5-Coder-32B-Instruct & 32B & 78.8 & 84.0 & 30.5 & 27.4 & 79.1 & 96.1 & 62.1 & \textbf{83.9} \\
        Qwen3-235B-A22B-Instruct-2507 & 22/235B & 62.0 & 89.5 & 53.9 & 51.8 & 80.4 & \textbf{96.9} & \textbf{62.8} & 81.1 \\
        Qwen3-235B-A22B-Thinking-2507 & 22/235B & 15.2 & 46.9 & 80.2 & 74.1 & 61.2 & 70.3 & 35.2 & 42.6 \\
        Qwen3-Coder-480B-A35B-Instruct & 35/480B & 87.1 & 90.4 & 48.6 & 53.9 & 80.2 & 96.1 & 61.3 & 81.2 \\
        Kimi-Dev-72B & 72B & 33.0 & 64.2 & 46.1 & 40.0 & 59.1 & 69.5 & - & - \\
        Kimi-K2-Instruct-0905 & 32B/1T & 86.8 & 89.5 & 52.1 & 53.7 & 76.1 & 90.6 & 60.4 & 81.1 \\
        Kimi-K2-Thinking & 32B/1T & \textbf{92.2} & 86.2 & - & 83.1 & 73.0 & 85.2 & 40.6 & 49.6 \\
        KAT-Dev & 32B & 42.5 & 65.1 & 32.9 & 32.6 & 75.1 & 89.1 & 52.2 & 77.6 \\
        KAT-Dev-72B-Exp & 72B & 71.4 & 81.1 & 13.8 & 16.0 & 79.0 & 94.5 & 35.2 & 60.3 \\
        GLM-4.7 & 32/355B & 65.6 & 81.2 & - & \textbf{84.9} & 74.1 & 86.7 & 46.5 & 62.4 \\
        InCoder-32B & 32B & 62.4 & 73.9 & 53.3 & 49.1 & 71.4 & 85.6 & 55.4 & 79.7 \\
        \rowcolor{tablegray} \textbf{\modelname{}} & 32B & 88.9 & \textbf{95.5} & \textbf{81.3} & 77.1 & 62.4 & 73.4 & 47.9 & 66.7 \\
    \bottomrule
    \end{tabular}
    }
\end{table*}

\begin{table*}[h]
    \centering
    \caption{Performance comparison on agentic coding tasks (Terminal-Bench v1.0, Terminal-Bench v2.0, SWE-Verified) and general tool-use tasks (Mind2Web, BFCL V3, $\tau^2$-bench).}
    \label{tab:agentic_combined}
    \small
    \setlength{\tabcolsep}{3.2pt}
    \renewcommand{\arraystretch}{1.05}
    \resizebox{1.0\textwidth}{!}{
    \begin{tabular}{lr|ccc|ccccc}
    \toprule
        \multirow{3}{*}{\textbf{Model}} & \multirow{3}{*}{\textbf{Size}} &
        \multicolumn{3}{c|}{\textbf{Agentic Coding}} &
        \multicolumn{5}{c}{\textbf{General Tool Use}} \\
        \cmidrule(lr){3-5} \cmidrule(lr){6-10}
        & & \multicolumn{2}{c}{\textbf{Terminal-Bench}} & \multirow{2}{*}{\textbf{\shortstack{SWE-bench\\Verified}}}
        & \multirow{2}{*}{\textbf{Mind2Web}} & \multirow{2}{*}{\textbf{BFCL V3}} & \multicolumn{3}{c}{\textbf{$\tau^2$-bench}} \\
        & & \textbf{v1.0} & \textbf{v2.0} & & & & \textbf{Airline} & \textbf{Retail} & \textbf{Telecom} \\
    \midrule

        \multicolumn{10}{c}{\textbf{6B+ Models}} \\
    \midrule
        DeepSeek-Coder-V2-Lite-Instruct & 2.4/16B & 5.0 &- & -   & 26.7 & - & 3.5 & 12.0 & - \\
        Qwen2.5-Coder-7B-Instruct       & 7B & 6.3 &- & -   & 38.4 & 54.2 & - & - & - \\
        Seed-Coder-8B-Instruct          & 8B & 7.5 & \textbf{2.5} & -   & 38.2 & - & \textbf{4.3} & \textbf{32.0} & - \\
        Qwen2.5-Coder-14B-Instruct      & 14B & \textbf{8.8}  &-  & -    & \textbf{42.7} & \textbf{59.9} & - & - & - \\
    \midrule
        \multicolumn{10}{c}{\textbf{30B+ Models}} \\
    \midrule
        Qwen3-Coder-30B-A3B-Instruct    & 3.3/30.5B & 23.8 & 23.8 & 51.9 & 36.1 & 63.4 & 42.0 & 25.4 & 25.4 \\
        DeepSeek-v3.2                   & 37/671B & 23.8 & \textbf{46.4} & 73.1 & 47.2 & 68.8 & 63.8 & 81.1 & \textbf{96.2} \\
        Qwen2.5-Coder-32B-Instruct      & 32B & 5.0  & 4.5  & -    & 32.5 & 62.3 & - & - & - \\
        Qwen3-235B-A22B-Instruct-2507   & 22/235B & 15.0 & 13.5 & 45.2 & 49.0 & 71.2 & 50.0 & 74.6 & 32.5 \\
        Qwen3-235B-A22B-Thinking-2507   & 22/235B & 8.8  & 3.4  & 44.6 & 43.2 & \textbf{71.9} & 58.0 & 71.9 & 45.6 \\
        Qwen3-Coder-480B-A35B-Instruct  & 35/480B & 37.5 & 23.6 & 67.0 & 54.0 & 68.7 & 60.0 & 77.5 & 65.8 \\
        Kimi-Dev-72B                    & 72B & -    & 2.3  & 60.4 & -    & 55.5 & 21.9 & 32.0 & 35.1 \\
        Kimi-K2-Instruct-0905           & 32B/1T & 44.5 & 27.8 & 69.2 & 53.4 & 70.3 & 56.5 & 70.6 & 65.8 \\
        Kimi-K2-Thinking                & 32B/1T & \textbf{47.1} & 33.7 & 71.3 & 55.7 & - & - & - & - \\
        KAT-Dev                         & 32B & 17.5 & 10.1 & 62.4 & 33.7 & 64.7 & 32.0 & 28.0 & 35.1 \\
        KAT-Dev-72B-Exp                 & 72B & 21.3 & 7.9  & 74.6 & -    & - & - & - & - \\
        GLM-4.7                         & 32/355B & 36.3 & 41.0 & 73.8 & 53.7 & 64.8 & 60.0 & 70.2 & 75.4 \\
        InCoder-32B     & 32B & 35.0 & 22.5 & \textbf{74.8} & \textbf{55.8} & 61.0 & \textbf{70.0} & 85.1 & 86.8 \\
        \rowcolor{tablegray}\textbf{\modelname{}}  & 32B & 38.8 & 21.6 & 70.4 & 49.1 & 63.9 & 62.0 & \textbf{86.0} & 95.6 \\
    \bottomrule
    \end{tabular}
    }
\end{table*}


\definecolor{tablegray}{gray}{0.92}
\begin{table*}[!t]
    \centering
    \caption{Performance comparison on chip design benchmarks. \modelname{} results are highlighted in gray.}
    \label{tab:chip_design_benchmarks}
    \resizebox{1.0\textwidth}{!}{
    \begin{tabular}{lr|c|c|cccccc|cc}
    \toprule
        \multirow{3}{*}{\textbf{Model}}
        & \multirow{3}{*}{\textbf{Size}}
        & \textbf{VeriScope}
        & \textbf{VeriRepair}
        & \multicolumn{6}{c|}{\textbf{RealBench}}
        & \multicolumn{2}{c}{\textbf{ArchXBench}} \\
        \cmidrule(lr){3-3} \cmidrule(lr){4-4} \cmidrule(lr){5-10} \cmidrule(lr){11-12}
        ~
        & ~
        & \multirow{2}{*}{\textbf{Score}}
        & \multirow{2}{*}{\textbf{Fix (\%)}}
        & \multicolumn{2}{c}{\textbf{System}}
        & \multicolumn{4}{c|}{\textbf{Module}}
        & \multirow{2}{*}{\textbf{$n$}}
        & \multirow{2}{*}{\textbf{$t$}} \\
        \cmidrule(lr){5-6} \cmidrule(lr){7-10}
        ~
        & ~
        & ~
        & ~
        & \textbf{Syn@1}
        & \textbf{Syn@5}
        & \textbf{Syn@1}
        & \textbf{Syn@5}
        & \textbf{Func@1}
        & \textbf{Func@5}
        & ~
        & ~ \\
        \midrule
        \multicolumn{12}{c}{\textbf{6B+ Models}} \\
        \midrule
        Qwen3.5-9B & 9B & 32.0 &- &- &- & 6.3 & 15.8 & 4.3 & 8.5 & 1.9 & 44.3 \\
        GPT-OSS-20B & 3.6/21B & \textbf{73.9} & \textbf{86.7} & 3.8 & 17.4 & \textbf{22.9} & \textbf{47.9} & 9.8 & \textbf{21.0} & \textbf{3.1} & \textbf{53.5} \\
        Qwen3.5-27B & 27B & 55.7 & 60.0 & \textbf{6.2} & \textbf{20.1} & 17.1 & 33.8 & \textbf{10.6} & 17.8 & 2.6 & 48.3 \\
        \midrule
        \multicolumn{12}{c}{\textbf{30B+ Models}} \\
        \midrule
        Qwen3-Coder-30B-A3B-Instruct & 3.3/30.5B & 66.2 & 76.7 &- &- & 23.0 & 35.2 & 5.2 & 8.2 & 2.4 & 37.3 \\
        Seed-OSS-36B-Instruct & 36B & 67.2 & 66.7 & 5.0 & 21.3 & 14.3 & 23.0 & 11.5 & 20.3 & 2.5 & 43.7 \\
        GPT-OSS-120B & 5.1/117B & 82.2 & 76.7 & 5.0 & 21.3 & 37.8 & 64.3 & 17.5 & 30.8 & 3.4 & \textbf{54.8} \\
        MiniMax-M2.5 & 10/230B & 75.1 & 66.7 & \textbf{23.8} & 48.5 & 17.2 & 38.4 & 6.9 & 15.9 & 2.9 & 46.0 \\
        GLM-4.7 & 32/355B & 81.2 & 63.3 & 12.5 & 24.6 & 25.4 & 46.9 & 11.6 & 21.2 & 3.2 & 51.4 \\
        GLM-5 & 40/744B & \textbf{83.2} & \textbf{90.0} & 2.5 & 11.2 & 22.2 & 43.4 & 12.2 & 22.6 & 3.1 & 53.2 \\
        Kimi-K2.5 & 32B/1T & 73.1 & 83.3 & 5.0 & 17.9 & 43.7 & 52.2 & 23.1 & 25.7 & \textbf{3.8} & 49.7 \\
        Kimi-K2-Instruct & 32B/1T & 82.4 & 76.7 & 6.2 & 26.2 & 50.1 & 70.1 & 22.2 & 28.3 & 2.9 & 44.9 \\
        Kimi-K2-Thinking & 32B/1T & 73.1 & 80.0 &- &- & 27.8 & 59.4 & 14.1 & 28.9 & 1.5 & 30.1 \\
        DeepSeek-V3.2 & 37/671B & 76.1 & 77.0 & 18.8 & \textbf{55.1} & 39.3 & 52.7 & 17.2 & 21.4 & 3.6 & 53.9 \\
        Qwen3.5-397B-A17B & 17/397B & 62.5 & 86.7 & 11.2 & 38.1 & 35.2 & 59.5 & 16.4 & 28.3 & 3.1 & 53.5 \\
        Qwen3-Coder-480B-A35B-Instruct & 35/480B & 80.8 & 76.7 &- &- & 28.9 & 39.5 & 14.8 & 20.6 & 3.0 & 43.9 \\
        InCoder-32B & 32B & 80.7 & 80.0 & 10.0 & 23.7 & 74.8 & \textbf{83.3} & 62.7 & \textbf{70.5} & 3.4 & 51.0 \\
        \rowcolor{tablegray} \textbf{\modelname{}} & 32B & 75.4 & 83.3 & 12.3 & 24.53 & \textbf{75.6} & 82.4 & \textbf{63.1} & 69.8 & 3.12 & 46.7 \\
        \midrule
        \multicolumn{12}{c}{\textbf{Closed-APIs Models}} \\
        \midrule
        Claude-Sonnet-4.6 & \faLock{} & 87.7 & 83.3 & 41.2 & 50.0 & 69.2 & 77.7 & 33.5 & 37.2 & 4.4 & 58.2 \\
    \bottomrule
    \end{tabular}
    }
\end{table*}



\begin{table*}[!t]
    \centering
    \caption{Performance on GPU optimization, code optimization, and 3D modeling benchmarks. \modelname{} results are highlighted in gray.}
    \label{tab:other_industrial_benchmarks}
    \resizebox{1.0\textwidth}{!}{
    \begin{tabular}{lr|cc|c|cc|cccc|ccc}
    \toprule
        \multirow{2}{*}{\textbf{Model}}
        & \multirow{2}{*}{\textbf{Size}}
        & \multicolumn{2}{c|}{\textbf{CAD-Coder}}
        & \textbf{EmbedCGen}
        & \multicolumn{2}{c|}{\textbf{SuperCoder}}
        & \multicolumn{4}{c|}{\textbf{TritonBench}}
        & \multicolumn{3}{c}{\textbf{KernelBench}} \\
        \cmidrule(lr){3-4} \cmidrule(lr){5-5} \cmidrule(lr){6-7} \cmidrule(lr){8-11} \cmidrule(lr){12-14}
        ~
        & ~
        & \textbf{Comp.}
        & \textbf{IoU}
        & \textbf{Main (\%)}
        & \textbf{Acc. (\%)}
        & \textbf{Spd.}
        & \textbf{G-call (\%)}
        & \textbf{G-exe (\%)}
        & \textbf{T-call (\%)}
        & \textbf{T-exe (\%)}
        & \textbf{L1}
        & \textbf{L2}
        & \textbf{L3} \\
        \midrule
        \multicolumn{14}{c}{\textbf{6B+ Models}} \\
        \midrule
        Qwen3.5-9B & 9B & 2.0 & - & 10.0 & \textbf{36.0} & 1.0$\times$ & 2.7 & \textbf{100.0} & 3.6 & \textbf{100.0} & - & - & - \\
        GPT-OSS-20B & 3.6/21B & 2.0 & \textbf{2.0} & \textbf{30.6} & 16.0 & 1.0$\times$ & 2.2 & \textbf{100.0} & 1.2 & \textbf{100.0} & \textbf{5.1} & \textbf{10.0} & 2.0 \\
        Qwen3.5-27B & 27B & \textbf{4.0} & 0.3 & 9.6 & 6.0 & \textbf{1.9}$\times$ & \textbf{5.4} & \textbf{100.0} & \textbf{25.9} & 97.7 & \textbf{5.1} & 6.0 & \textbf{4.0} \\
        \midrule
        \multicolumn{14}{c}{\textbf{30B+ Models}} \\
        \midrule
        Qwen3-Coder-30B-A3B-Instruct & 3.3/30.5B & - & - & 15.4 & 50.0 & 1.0$\times$ & 8.7 & \textbf{100.0} & 24.1 & 67.5 & - & - & - \\
        Seed-OSS-36B-Instruct & 36B & 2.0 & 2.0 & 10.2 & 8.0 & 1.1$\times$ & 1.6 & \textbf{100.0} & 1.8 & \textbf{100.0} & 1.0 & 2.0 & 2.0 \\
        GPT-OSS-120B & 5.1/117B & 4.0 & 1.9 & 17.8 & 8.0 & 1.2$\times$ & 3.8 & 85.7 & 12.7 & 95.2 & 6.1 & 15.0 & 2.0 \\
        MiniMax-M2.5 & 10/230B & 4.0 & 0.4 & 22.2 & 20.0 & 3.5$\times$ & 5.4 & \textbf{100.0} & 15.1 & \textbf{100.0} & 7.1 & 14.0 & 8.0 \\
        GLM-4.7 & 32/355B & 12.0 & 6.0 & 89.6 & 20.0 & \textbf{8.6}$\times$ & 3.3 & \textbf{100.0} & 6.0 & \textbf{100.0} & 8.1 & 19.0 &- \\
        GLM-5 & 40/744B & 38.0 & 18.8 & \textbf{90.2} & 54.0 & 1.87$\times$ & 1.6 & \textbf{100.0} & 1.2 & \textbf{100.0} & 16.2 & 23.0 & 4.0 \\
        Kimi-K2.5 & 32B/1T & 40.0 & 12.1 & 81.0 & 46.0 & 1.9$\times$ & 12.5 & \textbf{100.0} & 7.8 & \textbf{100.0} & 13.1 & 23.0 & 6.0 \\
        Kimi-K2-Instruct & 32B/1T & 2.0 & 1.1 & 69.6 & 12.0 & 1.1$\times$ & 15.8 & 96.5 & 13.9 & 91.3 & 6.1 &- &- \\
        Kimi-K2-Thinking & 32B/1T & 48.0 & 20.0 & - & 24.0 & 1.2$\times$ & 17.4 & \textbf{100.0} & 19.9 & 84.8 & 9.1 & 16.0 & 4.0 \\
        DeepSeek-V3.2 & 37/671B & 14.0 & 4.6 & 84.4 & 30.0 & 1.8$\times$ & 19.6 & \textbf{100.0} & 18.1 & 13.3 & 3.0 &- &- \\
        Qwen3.5-397B-A17B & 17/397B & 36.0 & 14.2 & 17.8 & 34.0 & 1.2$\times$ & 7.6 & \textbf{100.0} & 16.3 & 92.6 & 4.0 & 10.0 &- \\
        Qwen3-Coder-480B-A35B-Instruct & 35/480B & 10.0 & 4.7 & 9.0 & 64.0 & 2.0$\times$ & \textbf{20.1} & \textbf{100.0} & \textbf{31.9} & 56.6 & 3.0 & 6.0 &- \\
        InCoder-32B & 32B & 82.0 & \textbf{53.5} & 35.2 & 91.0 & 1.3$\times$ & 18.5 & \textbf{100.0} & 19.3 & 93.8 & \textbf{22.2} & 36.0 & \textbf{14.0} \\
        \rowcolor{tablegray} \textbf{\modelname{}} & 32B & \textbf{84.0} & 48.6 & 47.9 & \textbf{93.0} &  3.93$\times$& 15.2 & \textbf{100.0} & 22.9 & 89.5 & 20.2 & \textbf{38.0} & 12.0 \\
        \midrule
        \multicolumn{14}{c}{\textbf{Closed-APIs Models}} \\
        \midrule
        Claude-Sonnet-4.6 & \faLock{} & 77.0 & 32.4 & 79.0 & 88.0 & 4.6$\times$ & 28.8 & 98.1 & 41.6 & 1.4 & 11.1 & 28.0 & 2.0 \\
    \bottomrule
    \end{tabular}
    }
\end{table*}


\section{Evaluation}\label{sec:evaluation}
\subsection{Baselines}
We evaluate \modelname{} against a broad set of contemporary coding models, covering both open-weight systems and proprietary APIs. Because general-code and industrial-code benchmarks emphasize different capability profiles, we adopt separate baseline sets for the two evaluation settings and include strong reference models that are well matched to each benchmark family.

For general-code benchmarks, we compare \modelname{} against the models reported in \autoref{tab:code_generation_1}, \autoref{tab:code_reasoning_efficiency_sql}, and \autoref{tab:agentic_combined}. These baselines include DeepSeek-Coder-V2-Lite-Instruct~\cite{deepseek2024coder} and DeepSeek-V3.2~\cite{liu2025deepseekv32}; Qwen2.5-Coder-7B-Instruct, Qwen2.5-Coder-14B-Instruct, and Qwen2.5-Coder-32B-Instruct~\cite{qwen25coder}; Qwen3-235B-A22B-Instruct-2507 and Qwen3-235B-A22B-Thinking-2507~\cite{yang2025qwen3}; Qwen3-Coder-30B-A3B-Instruct and Qwen3-Coder-480B-A35B-Instruct~\cite{qwen3coder}; Seed-Coder-8B-Instruct~\cite{seedcoder}; Kimi-Dev-72B~\cite{kimi_dev}; Kimi-K2-Instruct-0905 and Kimi-K2-Thinking~\cite{team2025kimi2}; KAT-Dev and KAT-Dev-72B-Exp~\cite{katcoder}; and GLM-4.7~\cite{glm47}. Collectively, these baselines span a broad range of model scales and include both dense and mixture-of-experts architectures, providing a strong reference set for evaluating general coding ability.

For industrial-code benchmarks, we use the models reported in \autoref{tab:chip_design_benchmarks} and \autoref{tab:other_industrial_benchmarks}. The compared baselines include Qwen3.5-9B, Qwen3.5-27B, and Qwen3.5-397B-A17B~\cite{qwen3.5}; Qwen3-Coder families (30B-A3B and 480B-A35B)~\cite{qwen3coder,qwen_qwen3_coder_next_tech_report}; DeepSeek-V3.2~\cite{liu2025deepseekv32}; GLM-4.7~\cite{glm47} and GLM-5~\cite{zeng2026glm}; Kimi-K2.5~\cite{team2026kimi25}, Kimi-K2-Instruct~\cite{team2025kimi2}, and Kimi-K2-Thinking~\cite{team2025kimi2}; MiniMax-M2.5~\cite{minimax-m25}; Seed-OSS-36B-Instruct~\cite{seed2025seed-oss}; and GPT-OSS-20B and GPT-OSS-120B~\cite{agarwal2025gpt}. We also report Claude-Sonnet-4.6~\cite{claude46} as a proprietary reference model. This comparison set is chosen to cover leading models that are relevant to hardware-aware code generation, program optimization, and other domain-specific engineering workloads.

\subsection{Benchmarks}

\autoref{fig:benchmark_compare} provides a high-level overview of the benchmark landscape used in our evaluation, covering both general-code and industrial-code benchmarks. In the following, we describe these two benchmark families in more detail.

\paragraph{General Code Benchmarks.}
Following the evaluation setup of InCoder-32B~\cite{yang2026incoder32bcodefoundationmodel}, we assess \modelname{} on a broad suite of general code benchmarks spanning code generation, reasoning, efficiency, text-to-SQL, agentic coding, and tool use. The benchmark suite includes EvalPlus~\cite{evalplus} with HumanEval~\citep{chen2021codex} and MBPP~\citep{austin2021mbpp}, BigCodeBench~\cite{zhuo2024bigcodebench}, FullStackBench~\cite{liu2024fullstackbench}, CRUXEval~\cite{gu2024cruxeval}, LiveCodeBench~\cite{jain2024livecodebench}, Mercury~\cite{du2024mercurycodeefficiencybenchmark}, Spider~\cite{2018spider}, BIRD~\cite{2023bird}, Terminal-Bench~\cite{tbench2025}, SWE-bench Verified~\cite{swebenchverified}, Mind2Web~\cite{deng2023mind2web}, BFCL V3~\cite{patil2025bfcl}, and $\tau^2$-bench~\cite{barres2025tau2benchevaluatingconversationalagents}. Detailed benchmark descriptions and evaluation protocols can be found in \autoref{app:benchmark_details}.

\paragraph{Industrial Code Benchmarks.}
Our main emphasis is on industrial code evaluation, and we also follow the benchmark suite introduced in InCoder-32B~\cite{yang2026incoder32bcodefoundationmodel} to cover hardware-oriented and other engineering-intensive programming tasks. Specially, we report results on VeriScope, RealBench~\cite{jin2025realbench}, ArchXBench~\cite{purini2025mlcad}, and VeriRepair for chip design; KernelBench~\cite{ouyang2025kernelbenchllmswriteefficient} and TritonBench~\cite{li2025tritonbenchbenchmarkinglargelanguage} for GPU kernel optimization; EmbedCGen and SuperCoder~\cite{wei2026supercoderassemblyprogramsuperoptimization} for low-level code optimization; and CAD-Coder~\cite{guan2025cadcodertexttocadgenerationchainofthought} for 3D modeling. These benchmarks are substantially more demanding than standard software tasks because they require models to satisfy compilation and simulation constraints, reason about hardware and performance trade-offs, and produce outputs whose correctness is validated by domain-specific verification pipelines. We therefore use them as the primary testbed for measuring whether \modelname{} can operate effectively in realistic industrial coding scenarios.

\subsection{Main Results}

\paragraph{Results on General Code Benchmarks.}
\autoref{tab:code_generation_1}, \autoref{tab:code_reasoning_efficiency_sql}, and \autoref{tab:agentic_combined} summarize the results. The most striking finding is the leap in code reasoning. \modelname{} scores \textbf{81.3} on LiveCodeBench V5, the highest among all open-weight models including those with an order of magnitude more parameters, confirming that extended thinking can compensate for model size. Across code generation and agentic tasks, the model also remains broadly competitive, ranking first on $\tau^2$-bench Retail, while a moderate trade-off appears on benchmarks that favor concise responses (e.g., Mercury, Text2SQL), a pattern shared by other thinking-augmented systems.

\paragraph{Results on Industrial Code Benchmarks.}
\autoref{tab:chip_design_benchmarks} and \autoref{tab:other_industrial_benchmarks} reveal that the reasoning gains from our thinking-augmented training transfer effectively to industrial coding scenarios. On RealBench module-level chip design tasks, \modelname{} achieves the best open-weight scores by a wide margin. On CAD-Coder and SuperCoder, it surpasses even the proprietary Claude-Sonnet-4.6. Beyond these highlights, the model remains competitive on VeriScope, VeriRepair, and KernelBench. The fact that improvements span chip design, GPU programming, 3D modeling, and embedded systems suggests that the acquired reasoning abilities are not domain-specific but generalize broadly to hardware-aware and systems-level programming.

\section{Analysis}

\subsection{ICWM Fidelity Analysis}
\label{subsec:icwm_validation}

A core assumption of our pipeline is that the \textsc{ICWM} can faithfully replace real execution backends during large-scale trajectory synthesis.
To validate this assumption, we hold out 2{,}000 execution turns per domain from $\mathcal{D}_{\text{real}}$ and compare \textsc{ICWM} predictions against ground-truth labels along two complementary axes:
\textbf{outcome prediction accuracy}, the fraction of individual turns where the predicted label matches the real backend, and \textbf{trajectory agreement}, the fraction of full multi-turn trajectories whose final pass/fail verdict is identical under real and \textsc{ICWM}-driven execution.

\begin{figure}[t]
\centering
\includegraphics[width=\columnwidth]{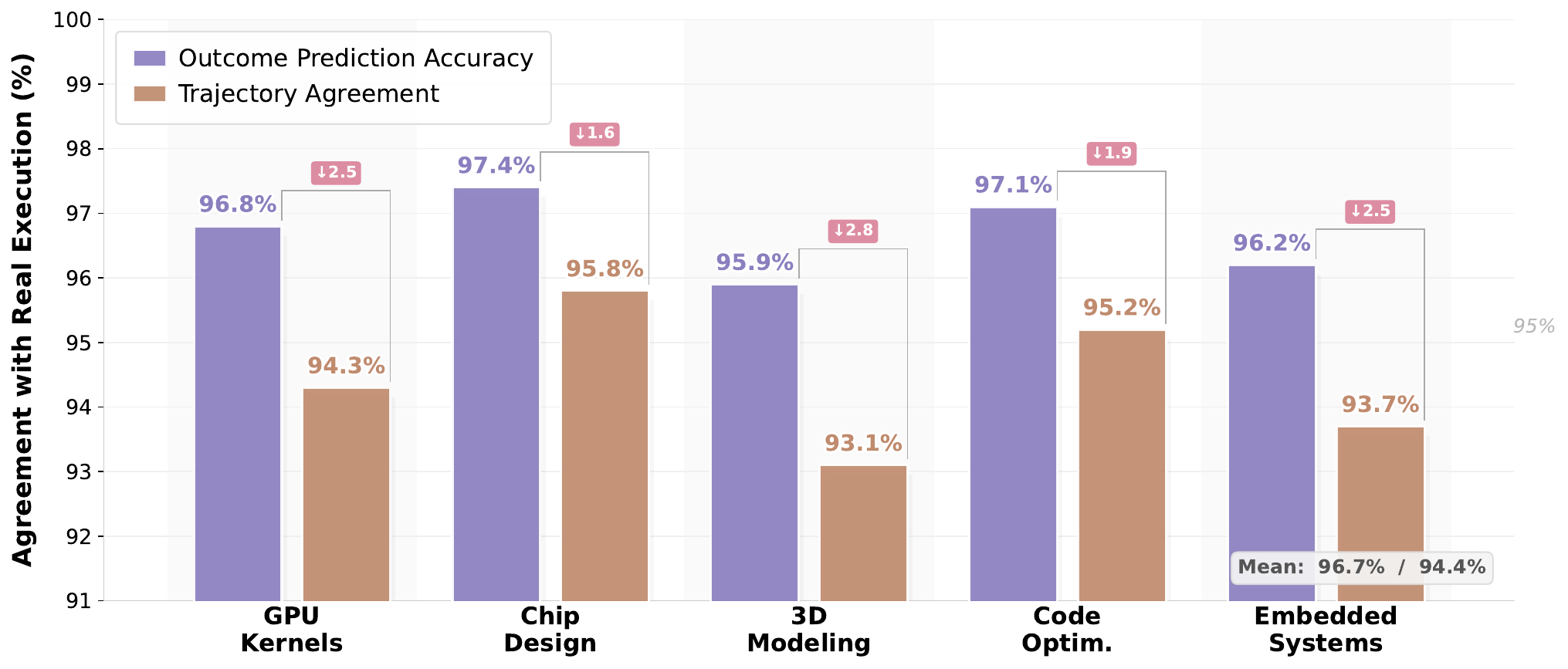}
\caption{ICWM fidelity across five industrial domains.
Outcome prediction accuracy measures per-turn label agreement; trajectory agreement measures end-to-end verdict consistency over multi-turn correction sequences.
The dashed line marks the 95\% threshold.
All domains exceed 95\% on outcome accuracy; the 1.6--2.8\,pp gap to trajectory agreement reflects error compounding across turns.}
\label{fig:icwm_validation}
\end{figure}

In \autoref{fig:icwm_validation}, the \textsc{ICWM} achieves a mean outcome prediction accuracy of 96.7\% and a mean trajectory agreement of 94.4\% across the five domains.
Chip design attains the highest fidelity at 97.4\%/95.8\%, as the Yosys and Icarus backends produce structurally template-like diagnostics that the world model learns reliably.
3D modeling presents the widest gap at 95.9\%/93.1\%, because CadQuery geometry checks involve floating-point tolerances and implicit Boolean operations whose outcomes are harder to predict from code text alone.
GPU kernels, code optimization, and embedded systems fall in between, with outcome accuracies of 96.2--97.1\%. The consistent 1.6$\sim$2.8 score drop from outcome accuracy to trajectory agreement is expected: a single mispredicted turn can redirect the generator onto a divergent correction path, compounding the initial error. Even at the lowest trajectory agreement of 93.1\%, more than nine out of ten \textsc{ICWM}-driven trajectories are functionally equivalent to their real-execution counterparts, confirming that the world model provides a reliable proxy for large-scale synthesis.

\paragraph{Case study: correct GPU kernel diagnosis.}
A Triton kernel for fused attention allocates a \texttt{tl.zeros([BLOCK\_M, BLOCK\_N])} tile in shared memory while launching with \texttt{num\_warps=8}, exceeding the SM's 48\,KB shared-memory budget.
The real Triton compiler returns \textsc{Memory\_Fault} with diagnostic \emph{``shared memory request (49152\,B) exceeds per-SM limit''}.
The \textsc{ICWM} predicts the same \textsc{Memory\_Fault} label with a quantitatively matching diagnostic message.
In the subsequent turn, the generator reduces \texttt{BLOCK\_N} from 128 to 64; the \textsc{ICWM} correctly predicts \textsc{Pass}.
This two-turn trajectory is identical under real execution, illustrating how the world model preserves the causal chain that drives error-correction reasoning.

\paragraph{Case study: 3D modeling false positive.}
A CadQuery script constructs a mounting bracket by subtracting a cylindrical bore from a rectangular body.
The candidate code places the cylinder axis exactly tangent to one face, creating a degenerate zero-thickness edge.
Real CadQuery raises \textsc{Geometry\_Error} with diagnostic \emph{``BRep check: edge has zero length''}.
The \textsc{ICWM} classifies this turn as \textsc{Pass} because the code is syntactically valid and the specified dimensions appear plausible, producing a false positive.
This type of geometric corner case accounts for the wider fidelity gap observed in 3D modeling.
During periodic audit rounds described in \autoref{subsec:icwm}, such mismatches are detected via real-execution spot checks, and the corrected labels are used to retrain the \textsc{ICWM}, progressively narrowing the gap.

\subsection{Adaptive Thinking Depth}
\label{sec:thinking_depth}
\begin{figure}[t]
\centering
\includegraphics[width=0.8\columnwidth]{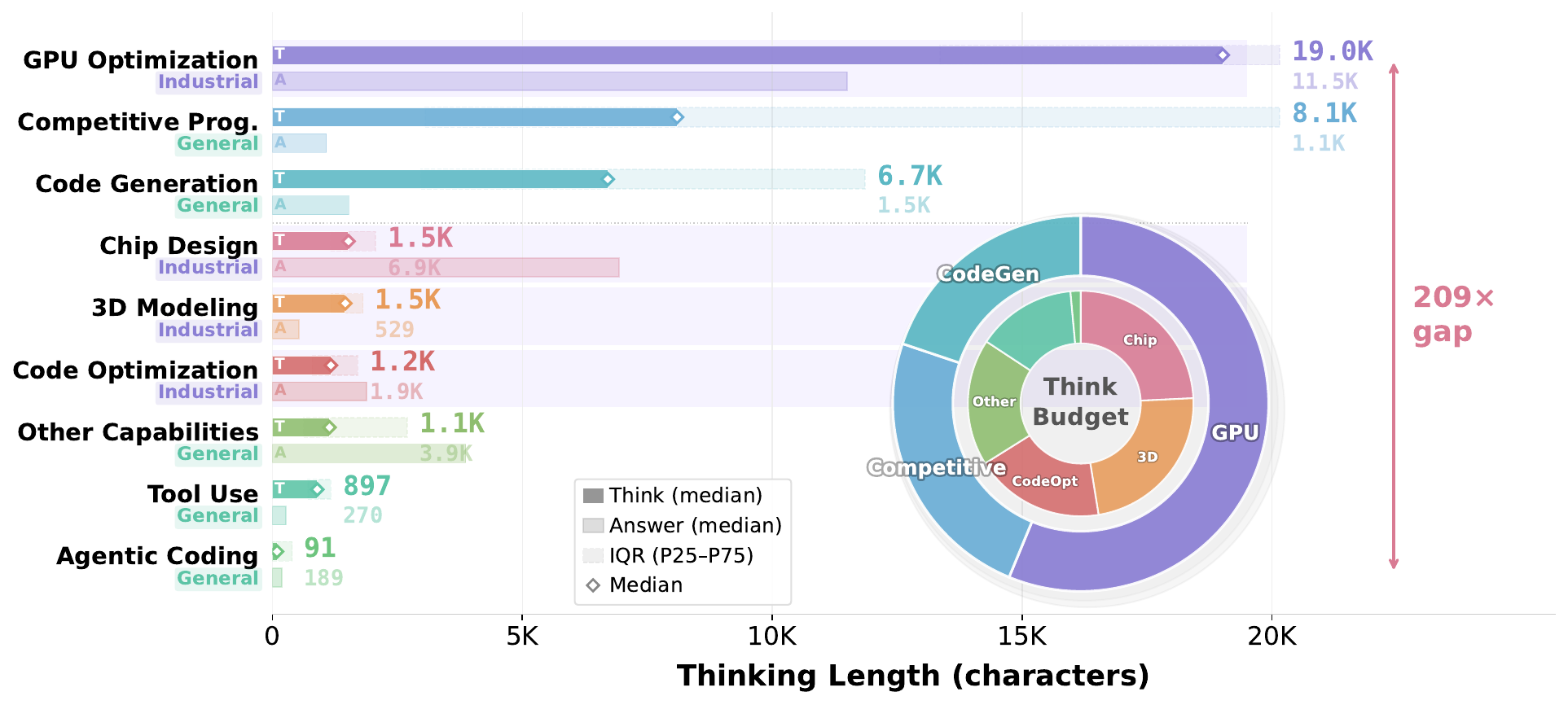}
\caption{Statistics of median thinking length (T) and answer length (A) per task category, sorted by thinking depth.
Shaded bands show the inter-quartile range (P25--P75).
Industrial domains are highlighted in purple.
Thinking depth spans a 209$\times$ range, reflecting the naturally varying reasoning demands of different execution backends.}
\label{fig:thinking_depth}
\end{figure}

\autoref{fig:thinking_depth} shows the distribution of \texttt{<think>} block lengths across the nine task categories in our training corpus.
The median thinking length spans a 209$\times$ range, from 91 characters per step in agentic coding to 19{,}015 characters in GPU kernel optimization.
This wide variation is not engineered by hand but arises naturally from the error-driven synthesis pipeline described in \autoref{sec:pipeline}: tasks involving complex execution feedback accumulate longer reasoning traces through successive correction rounds, whereas tasks with self-evident errors or multi-turn reasoning produce correspondingly shorter ones.

The domain-specific patterns closely mirror the nature of the real execution backends.
GPU kernel optimization demands the deepest reasoning, with a median of 19K characters, because each correction round requires diagnosing hardware-level issues such as grid/block configuration, shared-memory layout, and warp-level scheduling.
Chip design exhibits an inverted profile: a short thinking block of 1.5K characters followed by a long RTL answer of 6.9K, as the Yosys/Icarus feedback is structurally concise while the bulk of effort lies in code generation.
At the other extreme, agentic coding yields the shortest per-step thinking at 91 characters, since reasoning is distributed across tens of interaction turns and each step decides only the next action.

Because the thinking traces in $\mathcal{D}$ originate from real or ICWM-simulated execution feedback rather than a fixed prompting template, the training corpus naturally covers the full spectrum of reasoning depths.
The model therefore learns to adaptively calibrate its thinking effort, investing deep multi-step reasoning where complex industrial backends demand it while keeping traces succinct for tasks with short feedback loops.
This adaptive behaviour is a direct consequence of error-driven trajectory synthesis: the \texttt{<think>} blocks are shaped by the actual difficulty of each execution-feedback cycle, not by an externally imposed reasoning budget.

\begin{figure*}[t!]
    \centering
    \includegraphics[width=0.8\textwidth]{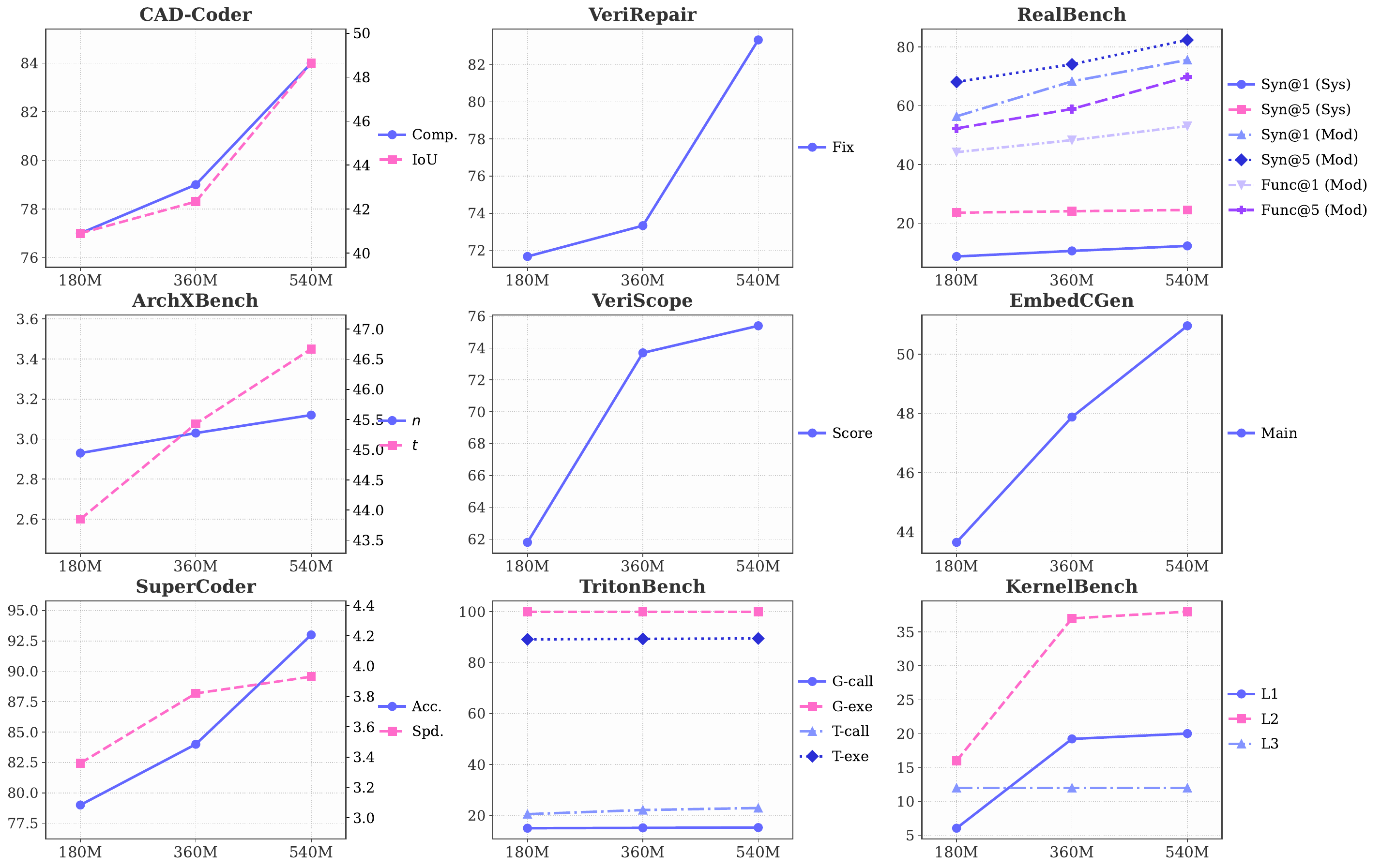}
    \caption{Performance visualization of \modelname{} across nine industrial code benchmarks as thinking training data scales from 180M to 540M tokens.}
    \label{fig:efftct_of_stage}
\end{figure*}

\subsection{Effects of Thinking Training Data}

Similar to the investigation of SFT data scale conducted for InCoder-32B, our objective is to explore the scaling impact of thinking training data. For this purpose, we collected a total of 540M tokens of thinking training data. This amount is much larger than the 250M tokens used for InCoder-32B. The reason for this difference is the chain of thought method. This reasoning process naturally creates longer texts and increases the total token count. To show the value of the thinking mechanism and how well the training data works, we tested \modelname{} checkpoints. These models were trained on 180M, 360M, and 540M tokens. As shown in \autoref{fig:efftct_of_stage}, most metrics improve steadily as the amount of thinking data grows. This proves that the thinking process is a strong driver for core industrial coding skills. With this mechanism, the VeriScope score jumps from 61.8 to 75.4, and the KernelBench L2 score rises from 16.0 to 38.0.

A few metrics stay flat, which also highlights the unique benefits of thinking data. For example, the TritonBench GPU execution correctness stays at a perfect 100 across all three stages. This shows that the thinking ability helps the model quickly master basic execution tasks. Similarly, the KernelBench L3 score remains at 12.0. This suggests that solving the hardest optimization problems requires specific strategies, not just more data volume. Overall, these results clearly show that adding more thinking training data is a highly reliable engine for unlocking deep reasoning skills.

\section{Related Work}

\subsection{Industrial Code Intelligence}
Applying LLMs to industrial software engineering presents distinct challenges compared to general-purpose programming, as these domains demand specialized syntax, hardware-aware reasoning, and rigorous functional correctness. Although significant progress has been made, existing approaches largely target isolated sub-domains.

In hardware design, early efforts fine-tuned general-purpose LLMs for RTL code generation~\cite{thakur2023verigen,liu2024rtlcoder}, and subsequent work extended the scope to Verilog debugging and formal verification~\cite{wang2025veridebug,yubeaton2025verithoughts}. Reinforcement learning with hardware-specific reward signals has further advanced this area, as demonstrated by CodeV-R1~\cite{zhu2025qimeng} and VeriReason~\cite{wang2025verireason}, while benchmarks such as VerilogEval~\cite{liu2023verilogeval} and RealBench~\cite{jin2025realbench} enable increasingly reliable evaluation. GPU kernel optimization represents another active frontier, where methods such as Kevin~\cite{baronio2025kevin} and CUDA Agent~\cite{dai2026cuda} leverage reinforcement learning for CUDA kernel synthesis and AscendKernelGen~\cite{cao2026ascendkernelgen} extends this paradigm to NPU targets. Other industrial domains, including embedded systems~\cite{yang2024embedgenius,xu2025embedagent}, compiler optimization~\cite{wei2025supercoder,cummins2025llm}, and 3D modeling~\cite{guan2025cad,li2025recad}, have also benefited from LLM-based approaches.

Despite these advances, existing work remains fragmented, with each model or benchmark tailored to a single industrial sub-domain. InCoder-32B~\cite{yang2026incoder32bcodefoundationmodel} represents a significant step toward unification by introducing the first 32-billion-parameter code foundation model that spans multiple industrial domains, including chip design, GPU kernel optimization, embedded systems, compiler optimization, and 3D modeling. Through a three-stage Code-Flow training pipeline comprising general code pre-training with industrial annealing, progressive context extension with synthetic industrial reasoning data, and execution-grounded post-training, InCoder-32B achieves competitive performance on general code benchmarks while establishing strong baselines across industrial domains. Building upon it, \modelname{} is a Thinking variant that integrates chain-of-thought reasoning to further enhance the model's capacity for complex industrial code generation.

\subsection{Thinking in Large Language Models}
OpenAI o1~\cite{openai_o1_2024} demonstrated that training models to produce long internal chains of thought via reinforcement learning (RL) can dramatically improve performance on complex reasoning tasks, establishing the concept of \emph{thinking models}. This line was advanced by o3~\cite{openai2025o3} and Gemini 3~\cite{gemini3pro}. On the open-source side, DeepSeek-R1~\cite{guo2025deepseek} showed that pure RL can incentivize emergent reasoning without supervised fine-tuning, while QwQ~\cite{qwq32b} and Qwen3~\cite{yang2025qwen3} introduced hybrid modes that dynamically switch between fast and slow thinking.

A central challenge for building thinking models lies in synthesizing high-quality long chains of thought. On the RL side, Group Relative Policy Optimization (GRPO)~\cite{shao2024deepseekmath} provides an efficient algorithm by removing the critic network required by PPO~\cite{ppo} and computing advantages over grouped samples. Beyond RL, reasoning distillation~\cite{deng2024explicit,opencode} offers a complementary pathway by transferring chain-of-thought capabilities from strong teacher models into smaller ones. STILL-2~\cite{min2024still2} proposed a three-stage pipeline of distillation, exploration via rejection sampling, and iterative self-improvement. A complementary line of work grounds reasoning in concrete execution signals rather than purely textual self-critique. Self-Refine~\cite{madaan2023selfrefine} introduced iterative refinement driven by model-generated feedback, and LeDex~\cite{jiang2024ledex} trained LLMs to self-debug code by learning from execution diagnostics. RLEF~\cite{gehring2024rlef} further grounded code LLMs in execution feedback via reinforcement learning, demonstrating that real execution outcomes can serve as a powerful reward signal.

For code-specific reasoning, structured chain-of-thought prompting~\cite{li2025structured}, o1-Coder~\cite{zhang2024o1}, and rStar-Coder~\cite{liu2025rstarcoder} have adapted thinking techniques to programming tasks, yet none target industrial code domains, nor do they leverage execution feedback to synthesize reasoning traces. \modelname{} unifies these research threads: it synthesizes long thinking content from multi-turn execution-grounded correction trajectories, where each failed attempt and its diagnostic feedback are folded into a coherent reasoning trace, and trains an industrial code world model to amplify reasoning data at scale without repeated access to real backends.

\section{Conclusion}
In this paper, we have presented \modelname{}, a thinking-augmented code model that bridges the gap between general code intelligence and the rigorous demands of industrial software development. By combining Error-driven chain-of-thought (ECoT) synthesis, which generates reasoning traces by explicitly modeling the iterative error-correction process, with an industrial code world model (ICWM) that learns causal dynamics between code and hardware behavior from domain-specific execution traces, \modelname{} acquires deep reasoning capabilities about hardware constraints without sacrificing general programming performance. Extensive evaluations across 14 general and 9 industrial benchmarks demonstrate that \modelname{} achieves 81.3\% on LiveCodeBench V5, while establishing the strongest open-source results across chip design, GPU optimization, embedded systems, and 3D modeling. Our analysis further reveals that scaling thinking data from 180M to 540M tokens yields consistent industrial improvements, confirming that chain-of-thought synthesis grounded in world model guidance is key to enabling reasoning in real-world engineering scenarios.

\clearpage
\newpage

\bibliography{ref}

\newpage
\appendix
\section{Benchmark Details}\label{app:benchmark_details}

\subsection{General Code Benchmarks}

\paragraph{Code Generation.}
\textbf{EvalPlus}~\cite{evalplus} augments HumanEval~\cite{chen2021codex} and MBPP~\cite{austin2021mbpp} with substantially more test cases, providing a more rigorous assessment of functional correctness for short Python programs.
\textbf{BigCodeBench}~\cite{zhuo2024bigcodebench} tests the ability to invoke complex library APIs through 1,140 task-oriented prompts that require composing calls across popular Python packages.
\textbf{FullStackBench}~\cite{liu2024fullstackbench} covers 16 programming languages and 4,000+ problems across multiple application domains, measuring full-stack development capability.

\paragraph{Code Reasoning.}
\textbf{CRUXEval}~\cite{gu2024cruxeval} probes code understanding via input and output prediction tasks on 800 short Python programs, requiring the model to trace program execution.
\textbf{LiveCodeBench}~\cite{jain2024livecodebench} collects competitive-programming problems from LeetCode, AtCoder, and Codeforces, with a contamination-free time-stamped split; we evaluate on V5 and V6.

\paragraph{Code Efficiency.}
\textbf{Mercury}~\cite{du2024mercurycodeefficiencybenchmark} evaluates whether models can generate runtime-efficient solutions, reporting Beyond@1 and Pass@1 to measure both efficiency and correctness.

\paragraph{Text-to-SQL.}
\textbf{Spider}~\cite{2018spider} and \textbf{BIRD}~\cite{2023bird} are cross-database text-to-SQL benchmarks that require translating natural-language questions into SQL queries, testing schema understanding and complex join/aggregation reasoning.

\paragraph{Agentic Coding.}
\textbf{Terminal-Bench}~\cite{tbench2025} evaluates multi-turn terminal interactions for software engineering tasks, with v1.0 and v2.0 presenting problems of increasing complexity.
\textbf{SWE-bench Verified}~\cite{swebenchverified} requires resolving real GitHub issues end-to-end, including code understanding, patch generation, and test validation.

\paragraph{Tool Use.}
\textbf{Mind2Web}~\cite{deng2023mind2web} benchmarks web navigation agents that must select correct elements and actions on real websites.
\textbf{BFCL V3}~\cite{patil2025bfcl} evaluates function calling accuracy across diverse API schemas.
\textbf{$\tau^2$-bench}~\cite{tau2bench} measures multi-turn conversational agent performance in three customer-service domains (Airline, Retail, Telecom), where the agent must resolve user requests through tool-augmented dialogue.

\subsection{Industrial Code Benchmarks}

\subsubsection{Chip Design}

\paragraph{VeriScope.}
We propose VeriScope, a Verilog generation benchmark containing 568 problems organized into five difficulty tiers. Tasks span basic combinational logic and hierarchical module composition at lower levels, progressing to full system-level designs, up to extreme challenges such as implementing a dual-core out-of-order RISC-V SoC with cache coherence at the highest tier. Correctness is assessed via simulation with a three-level scoring scheme: a submission that fails compilation receives 0 points, one that compiles but does not pass the functional testbench receives 50 points, and one that passes all test cases receives the full 100 points. We report the average score across all problems.

\paragraph{RealBench.}
RealBench~\cite{jin2025realbench} focuses on production-grade IP-level Verilog generation rather than small algorithmic snippets. The benchmark is constructed from four real-world open-source IP cores, including AES encryption, an SD card controller, and the Hummingbirdv2 E203 CPU. It offers 60 module-level subtasks, where missing sub-modules may fall back to golden reference implementations, and 4 system-level subtasks that demand implementing the entire module hierarchy from scratch given only a top-level specification. We generate 20 independent samples per task and adopt a layered verification pipeline with two families of metrics: Syn@$k$ measures whether at least one of $k$ candidates compiles without error, while Func@$k$ further checks functional correctness by running testbench simulation on the compilable candidates. We report both metrics at $k \in \{1, 5\}$ for system-level and module-level tasks separately.

\paragraph{ArchXBench.}
ArchXBench~\cite{purini2025mlcad} comprises 51 complex digital-system designs spanning six difficulty levels, with tasks drawn from domains such as cryptography, signal processing, image processing, and machine learning. Designs range from basic arithmetic circuits to full subsystems like AES cores and streaming FFT/DCT pipelines. Each task is accompanied by a problem description, an interface specification, and a testbench. We sample five independent candidates per task and report results in an $(n, t)$ format: $n$ is the number of syntactically correct candidates out of five, and $t$ is the percentage of testbench assertions passed by the best candidate among them.

\paragraph{VeriRepair.}
As part of this work, we construct VeriRepair, a Verilog error diagnosis and repair benchmark built by systematically injecting realistic bugs into verified implementations. The error taxonomy covers 4 major categories and 20 fine-grained error types, including syntax errors, type and structural errors, timing and FSM errors, and semantic/logic errors. The dataset consists of approximately 22,000 training samples and 300 test samples, each annotated with the buggy code, error category, error location, a corrected reference, and a testbench. We focus on the repair task and report the fix rate (Fix\%), defined as the fraction of buggy programs that the model successfully corrects, as verified by testbench execution.

\subsubsection{GPU Kernel Optimization}

\paragraph{KernelBench.}
KernelBench~\cite{ouyang2025kernelbenchllmswriteefficient} presents 250 PyTorch ML workloads organized into three levels of increasing complexity: Level~1 contains single operator invocations, Level~2 features operator sequences that are amenable to kernel fusion, and Level~3 consists of end-to-end model architectures. Given a PyTorch reference implementation, the model must produce a functionally equivalent but faster version using any available optimization tool. We adopt the $\text{fast}_p$ metric, which measures the fraction of tasks where the generated kernel is both correct and achieves a speedup of at least $p\times$ over the PyTorch baseline. Results are reported as $\text{fast}_1$ separately for L1, L2, and L3.

\paragraph{TritonBench.}
TritonBench~\cite{li2025tritonbenchbenchmarkinglargelanguage} is designed to evaluate Triton operator generation, targeting the Python-like GPU programming language widely adopted in production systems such as vLLM and Liger-kernel. The benchmark provides two evaluation tracks: TritonBench-G contains 184 curated real-world operators collected from GitHub repositories across five difficulty levels, while TritonBench-T aligns tasks with standard PyTorch operator interfaces. For each track we measure two metrics: call accuracy (whether the generated code can be executed without runtime error) and execution accuracy (whether the output further matches the reference implementation). This yields four reported metrics: G-call, G-exe, T-call, and T-exe.

\subsubsection{Code Optimization}

\paragraph{EmbedCGen.}
We also contribute EmbedCGen, a benchmark for bare-metal embedded C code generation targeting resource-constrained microcontrollers. It includes 500 problems across five difficulty tiers, starting from basic peripheral control and register-level operations and progressing to complex multi-peripheral integration, DMA mechanisms, and state-machine coordination. Generated code must adhere to HAL conventions and satisfy hard real-time constraints. Evaluation follows a strict serial pipeline: code generation, cross-compilation via the ARM GCC toolchain, and functional verification through the Renode system-level simulator. We report the average pass rate (Main\%) across all 500 problems.

\paragraph{SuperCoder.}
SuperCoder~\cite{wei2026supercoderassemblyprogramsuperoptimization} formulates assembly superoptimization as a language modeling task: given a C program together with its \texttt{gcc -O3} compiled output, the model must produce a semantically equivalent assembly program that runs faster. The benchmark comprises 8,072 x86-64 assembly programs averaging 130 lines each, with accompanying test suites that achieve 96.2\% line coverage and 87.3\% branch coverage. Two metrics are reported: accuracy (Acc.\%), which measures functional correctness through the test suite, and speedup (Spd.), which quantifies the runtime improvement relative to the compiler-optimized baseline.

\subsubsection{3D Modeling}

\paragraph{CAD-Coder.}
CAD-Coder~\cite{guan2025cadcodertexttocadgenerationchainofthought} casts text-to-CAD generation as producing executable CadQuery scripts, a Python-based parametric CAD language built on top of the OpenCascade geometry kernel. The benchmark is derived from the Text2CAD dataset~\cite{khan2024text2cad}, yielding 110K verified text--CadQuery--3D model triplets stratified by quality into 8K high-quality, 70K medium-quality, and 32K hard cases, along with 1.5K chain-of-thought annotated samples. Two evaluation metrics are employed: the compilation success rate (Comp.) measures the percentage of generated scripts that execute successfully and produce valid 3D geometry, while the IoU score quantifies geometric fidelity by computing the volumetric overlap between the generated voxelization and the ground-truth model.

\end{document}